\documentclass{optica-article}

\journal{opticajournal} 

\articletype{Research Article}

\begin{document}

\title{Viewing-angle expansion of holographic image using enhanced-NA Fresnel hologram}

\author{Byung Gyu Chae}

\address{Holographic Contents Research Laboratory, Electronics and Telecommunications Research Institute, 218 Gajeong-ro, Yuseong-gu, 
Daejeon 34129, Republic of Korea}

\email{bgchae@etri.re.kr}

\begin{abstract*}
The expansion of viewing angle is a crucial factor in holographic displays implemented with a spatial light modulator having a finite space-bandwidth.
The enhanced-NA Fresnel hologram reconstructs a holographic image at an angle larger than the diffraction angle by a hologram pixel,
where it has a difficulty in achieving this without an interference of high-order noises.
This study presents the theoretical foundation for optimizing the enhanced-NA Fresnel hologram to recover the low space-bandwidth.
The higher spectrum components of the digital hologram beyond the bandwidth exists in the form of their replications.
The expansion of angular spectrum by its repetition during optimization procedure increases the image resolution, 
resulting in a viewing angle that is dependent on the hologram numerical aperture.
We numerically and experimentally verify our strategy to expand a viewing angle of holographic image.

\end{abstract*}

\section{Introduction}

A holographic display is an optical imaging system that forms a three-dimensional image in free space \cite{1}.
The spatial resolution of a focused image is explained by the Abbe diffraction limit \cite{2,3,4,5}, 
where the lateral resolution represented as the hologram numerical aperture (hNA) has a direct connection with the angular field view of the holographic image.
When the hologram is recorded, the geometric structure, which is the ratio of the hologram aperture size to the distance,
specifies the range of viewing angle $\it{\Omega}$ \cite{6}:
\begin{equation}
\it{\Omega} = \rm{}2 \sin^{-1} (hNA).
\end{equation}
The hNA is given by the maximum spatial frequency $f_{\rm{max}}$ of the hologram fringe,
$\rm{hNA} = \it{} \lambda f_{\rm{max}}$, where $\it{\lambda}$ is the wavelength of the recoding wave.
In an analog hologram, this value can determine the viewing angle because nanoscale recording materials are usable.
In a holographic display, the digital hologram is encoded on a pixelated spatial light modulator (SLM) with a finite pixel pitch.
To realize a holographic image with a wide viewing-angle, the wavelength-scale SLM should be developed.
Several researchers have tried to tackle this problem by expanding the diffraction zone in terms of spatial and temporal multiplexing of the pixelated modulator \cite{7,8,9},
or the use of the spatial grid with a fine pitch \cite{10,11,12}.

In the strategy to expand the diffraction zone, it is natural for the upper bound of viewing angle to exist in a specific pixelated modulator.
However, it is not simple because of the sampling characteristics of the digital hologram.
When the sampling rate is larger than the bandwidth of the digital hologram, aliased errors occur according to the Nyquist sampling criterion \cite{13,14,15}.
This alias error is non-trivial unlike simple aliasing in conventional image processing.
We have reported that in numerical and optical experiments \cite{15,16}, the aliasing effect creates replica fringes of the original pattern,
contributing to the reconstruction of the original image with a high resolution.
That is, the replica fringes still play a role as components of the higher angular spectrum producing a propagating diffraction field into a wide angular region. 
This phenomenon is caused by the unusual properties of an optical kernel function with a quadratic phase.
We defined this type of digital hologram implying non-trivial aliasing errors as the enhanced-NA Fresnel hologram \cite{15}.

This kind of Fresnel hologram reconstructs a holographic image at an angle larger than the diffraction angle by a hologram pixel,
whereas the image space is confined to the diffraction zone due to the high-order terms.
High-order images result from the deviation of the sampling conditions in the hologram plane.
Thus, to reconstruct a holographic image at a wide angle while maintaining the image size, an algorithm for removing the aliasing noises should be developed.
Previously, we developed the algorithm for extending the image space of a holographic image beyond the diffraction zone \cite{17}, which is based on the single Fourier transfom.
However, it still has a limitation for acquring a sufficient viewing-angle dependent on the numerical aperture of digital hologram.

In this study,
we present the algorithm that adequately synthesizes the Fresnel hologram with a high numerical aperture through the double Fourier transform.
A crucial idea exists in using the expansion of angular spectrum of digital hologram in the Fourier space.
First, we describe a principle of the proposed algorithm, and examine the characteristics of the synthesized digital hologram and the reconstructed image.
Further, numerical and optical experiments are performed to confirm the proposed concept for expanding a viewing angle.

\begin{figure}[ht!]
\includegraphics[scale=0.7, trim=0.3cm 10.5cm 0cm 0cm]{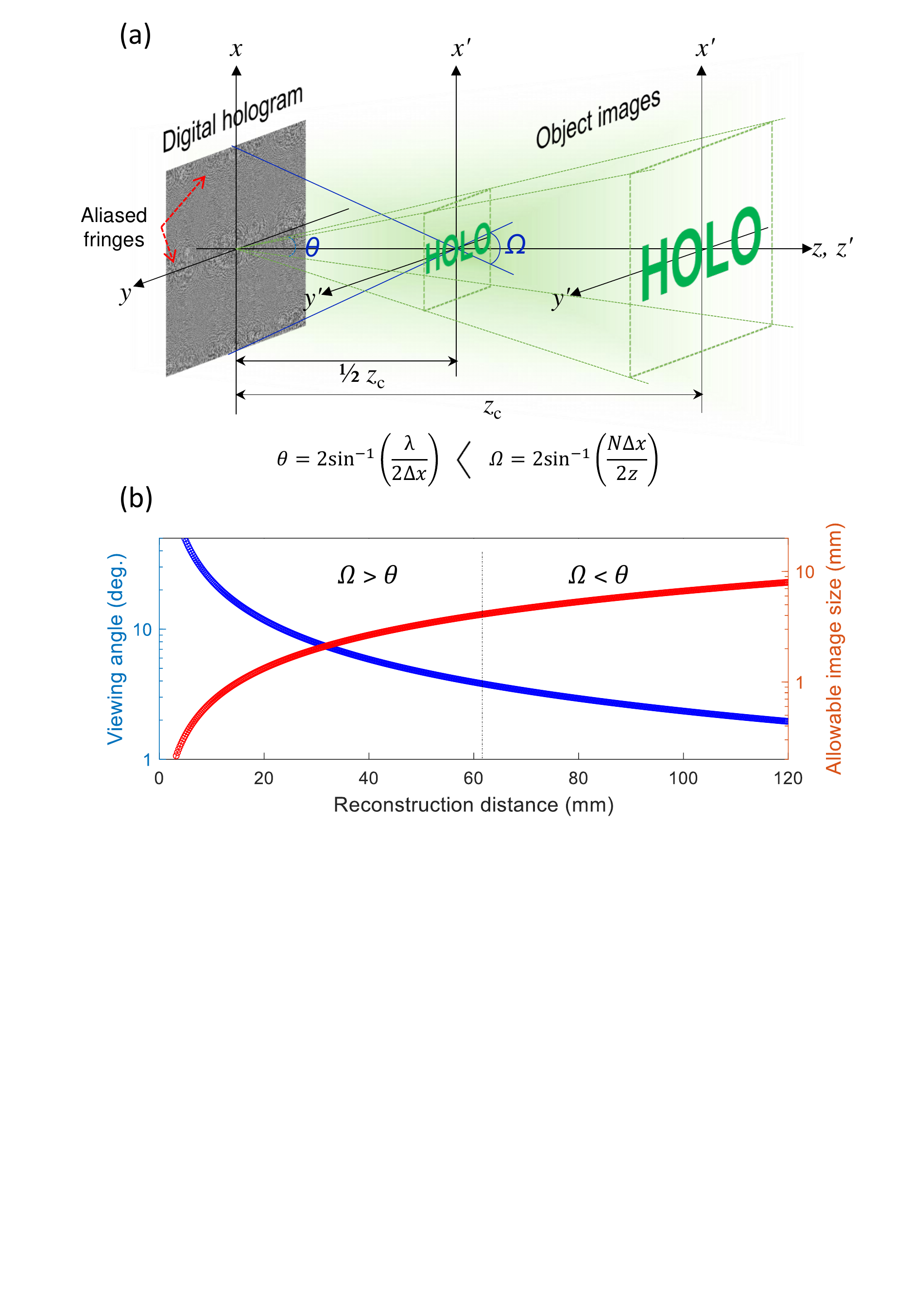}
\caption{Schematic of the enhanced-NA Fresnel holographic display system.
(b) Changes in the viewing angle $\it{\Omega}$ and allowable image size with a reconstruction distance. 
Diffraction angle $\theta$ for digital hologram consisting of 512$\times$512 pixels with an 8-$\mu$m pixel pitch is 3.8$^\circ$.}
\end{figure}

\section{Principle of expanding angular spectrum of digital hologram}
\subsection{Viewing angle of holographic image in enhanced-NA Fresnel hologram}

The hologram field $g(x,y)$ propagating between an object $o(x',y')$ and a hologram plane is obtained using the expanded Fresnel equation \cite{1}:
\begin{equation}
g(x,y) = \frac{e^{i kz}}{i\lambda z} \exp \left[\frac{i\pi}{\lambda z} (x^2 + y^2) \right]
\textbf{\textit{FT}} \left[ o(x',y') \exp \left[\frac{i\pi}{\lambda z} (x'^2 + y'^2) \right] \right],
\end{equation}
where $k$ is the wavenumber of $2\pi/\lambda$, $z$ is the propagation distance, 
and $\textbf{\textit{FT}}$ is the Fourier transform.
The sampling condition of the digital hologram can be interpreted from the calculated diffraction field in digitized space. 
The enhanced-NA Fresnel hologram is synthesized at a distance shorter than the critical distance $z_c$, as illustrated in Fig. 1(a).
The critical distance, $z_c = N \it{\Delta} x^{\rm{2}} /\lambda$
is defined by the sampling relation in both planes \cite{15,18},
\begin{equation}
\it{\Delta} x = \frac{\lambda z}{N\Delta x'}.
\end{equation}
We adopted a one-dimensional description for the sampling analysis,
where $\it{\Delta} x$ and $\it{\Delta} x'$ indicate the pixel pitches in the respective planes consisting of $\it{N}$$\times$$\it{N}$ pixels.
The lateral resolution of the object image is finer than the hologram pixel pitch.
We previously studied the dependency of the viewing angle of the reconstructed image on the hologram numerical aperture \cite{6},
\begin{equation}
\rm{hNA} = \rm{} \sin \left( \frac {\it{\Omega}} {\rm{2}} \right) = \frac{\it{N\Delta x}}{2\it{z}},
\end{equation}
which determines the image resolution.
Here, the viewing angle $\it{\Omega}$ is larger than the diffraction angle $\it{\theta}$ by a hologram pixel pitch,
\begin{equation} 
\it{\Omega} = \rm{}2 \sin^{-1} \left( \frac{\it{N\Delta x}}{2\it{z}} \right)  > \theta = \rm{}2 \sin^{-1} \left( \frac {\lambda}{2 \it{\Delta} x} \right).
\end{equation}
This value increases at a shorter distance because of the high hNA, in Fig. 1(b).
This type of hologram suffers from aliased errors owing to undersampling, but the original image can be completely retrieved. 
The Fresnel matrix undersampled by a regular interval still has a unitary property, which enables stable recovery under a severe aliasing error \cite{15}.

However, the image space is confined to the diffraction zone by a hologram pixel to avoid the interference of high-order diffractions, 
as shown in Fig. 1.
Algorithms consisting of the double Fourier transform are effective method for calculating the digital hologram using the object space beyond the diffraction scope:
\begin{equation}
g(x,y) = \textbf{\textit{IFT}} \left( \textbf{\textit{FT}} \left[ o(x',y') \right] \textbf{\textit{FT}} \left[ \exp \left[\frac{i\pi}{\lambda z} (x'^2 + y'^2)  \right] \right] \right).
\end{equation}
Figure 2 illustrates the digital holographic system based on a convolutional method.
In this algorithm, the hologram pixel pitch is calculated via the pixel quantity $\it{\Delta} f$ of the intermediate Fourier plane,
\begin{equation}
\it{\Delta} x = \frac{\rm{1}}{N\Delta f} = \it{\Delta} x',
\end{equation}
and thus, the sizes of the hologram and object are the same.
On the other hand, the high-frequency components of digital hologram are ruled out because only the zeroth-order region in the 
intermediate Fourier plane, $N\it{\Delta}f$ is used.

\begin{figure}[ht!]
\includegraphics[scale=0.7, trim= 0.5cm 19cm 0cm 0cm]{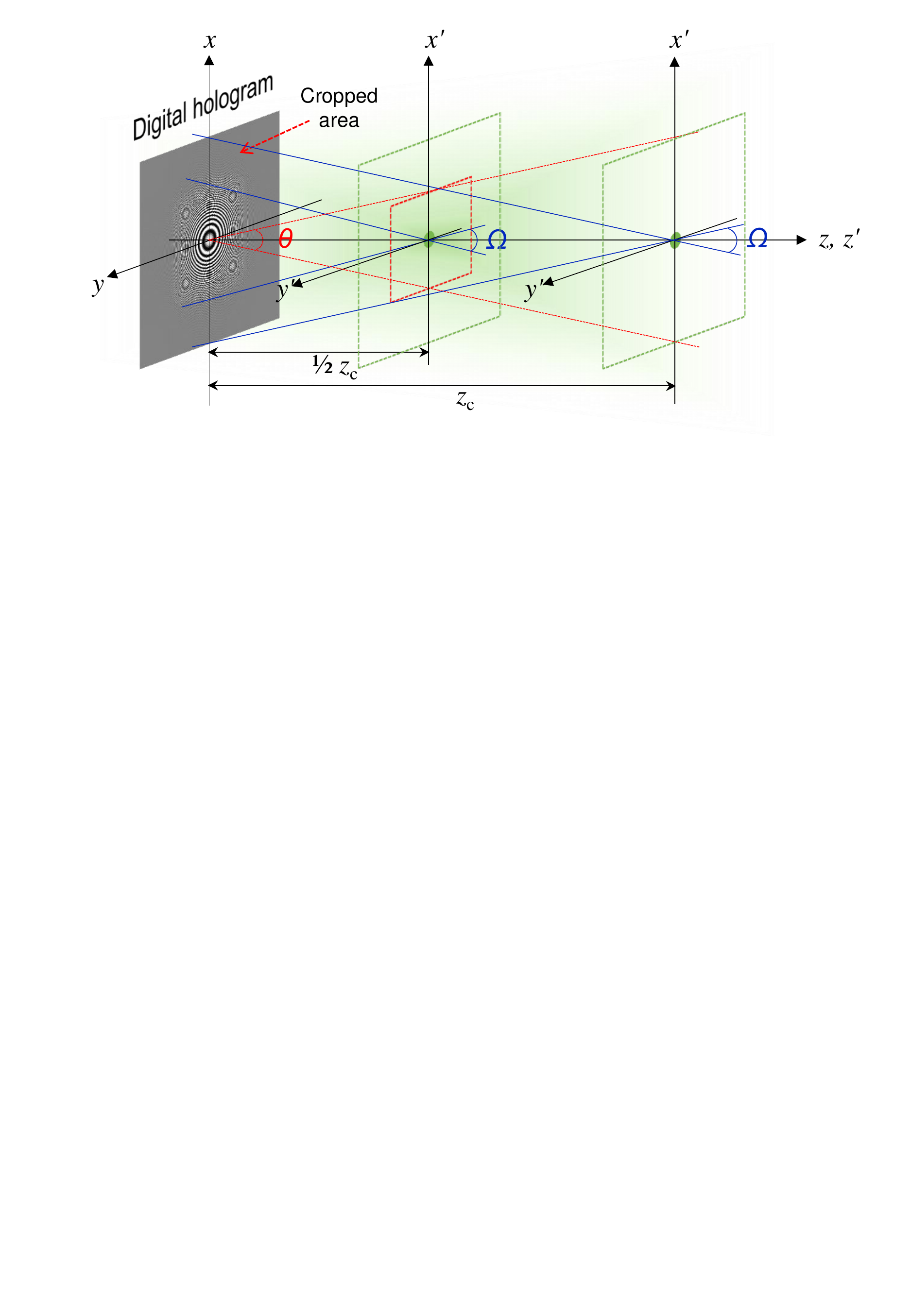}
\caption{Digital holographic system via the method consisting of double Fourier transform. The hologram for a point object is depicted for convenience.
Viewing angle $\it{\Omega}$ and diffraction angle $\theta$ are identical irrespective of reconstruction distance.}
\end{figure}

In the enhanced-NA Fresnel hologram, the high-frequency components are recorded as aliased replica fringes, as depicted in Fig. 1(a).
The aliased fringes are patterns undersampled by a hologram pixel pitch larger than the required value in the sampling condition.
The Fourier-transformed data does not include a higher angular spectrum of the hologram field.
This characteristic fundamentally prohibits aliased errors in both planes, as reported in a previous research \cite{6}.
By cropping the aliasing fringes in the algorithm, the hNA is maintained at a constant value irrespective of the distance, that is, the viewing angle does not vary.

\subsection{Expansion of angular spectrum by its replication in Fourier space}

To overcome the fixation of numerical aperture in the method implemented with the double Fourier transform, a higher angular spectrum should be utilized during the algorithm process.
Optical kernel functions in real and Fourier spaces play a role as a basis function in digital hologram synthesis.
We investigated the properties of the optical kernel functions when being sampled, that is, the impulse response function in the real space,
\begin{equation}
h(x,y) = \frac{e^{i kz}}{i\lambda z} \exp \left[\frac{i\pi}{\lambda z} (x^2 + y^2) \right]
\end{equation}
and the transfer function in the frequency domain,
\begin{equation}
H(f_x,f_y) = e^{i kz} \exp \left[-i\pi \lambda z (f_x^2 + f_y^2) \right].
\end{equation}
The response function is the convolutional kernel used to calculate the hologram field, where this function itself becomes a complex hologram for a point object $\delta (x,y)$.
Likewise, the transfer function is the angular spectrum of the point’s hologram. 
Digital hologram of an object with a finite size is characterized by the summation of individual functions.

\begin{figure}[ht!]
\centering\includegraphics[scale=0.65, trim= 0cm 11cm 0cm 0cm]{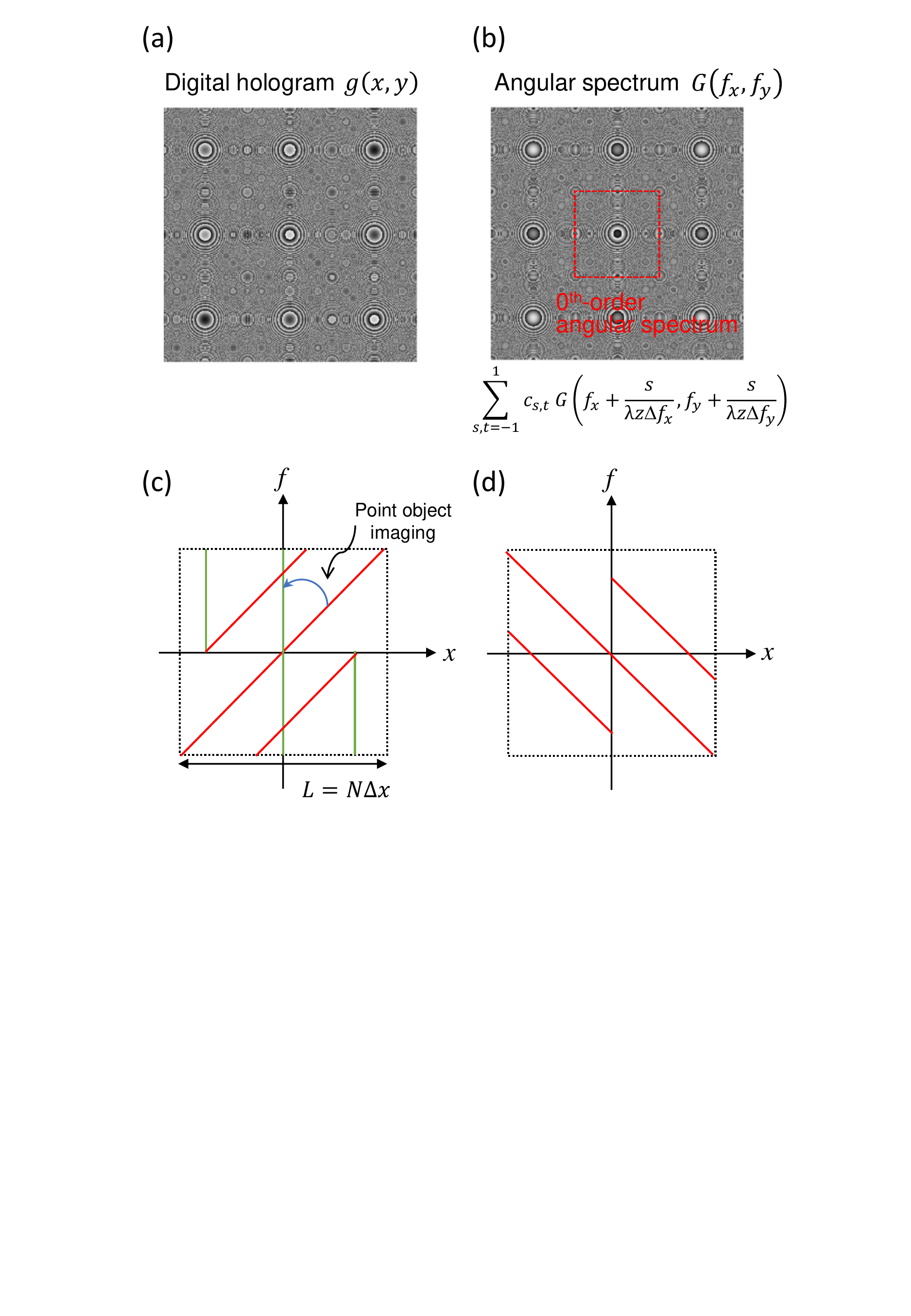}
\caption{Numerical analysis for describing the replication properties of optical kernel functions in real and Fourier spaces.
(a) Digital hologram for a point object synthesized at a distance of one-third of $z_c$. It is represented as an impulse response function.
(b) 3-fold expanded angular spectrum corresponding to digital hologram.
Wigner distribution description of (c) digital hologram and (d) angular spectrum for point object imaging. 
Point image is focused from inclined line designating spherical fringe of hologram.}
\end{figure}

The Fourier transform of sampled response function by a regular interval $\it{\Delta}x$ is given by
\begin{equation}
\textbf{\textit{FT}} \left[ \sum_{n} h(n\it{\Delta} x) \delta(x-n\it{\Delta} x) \right] = \frac{1}{\it{\Delta} x} \sum_{p} H \left( f - \frac{p}{\it{\Delta} x} \right).
\end{equation}
Taking the inverse Fourier transform into the right term of Eq. (10), we get the modulated form of this function,
which is rewritable in the shifted form of continuous quadratic functions \cite{16,19},
\begin{equation}
\sum_{n} h(n\it{\Delta} x) \delta(x-n\it{\Delta} x) = \frac{\rm{1}}{\it{\Delta} x} \sum_{n} c_n h \left( x + \frac{\lambda zn}{\it{\Delta} x} \right).
\end{equation}
The replica functions are formed at a reduced period of $\lambda z /(s\it{\Delta} x)$ when the function is undersampled by $s$ multiples of $\it{\Delta} x$.
High spatial-frequency components of hologram field are recorded in the form of replica fringes.
Similarly, when the transfer function is sampled, the same form as that in Eq. (11) is obtained:
\begin{equation}
\sum_{q} H(q\it{\Delta} f) \delta(f-q\it{\Delta} f) = \frac{\rm{1}}{\it{\Delta} f} \sum_{q} c_q H \left(f + \frac{q}{\it{\lambda z\Delta} f} \right).
\end{equation}
The higher angular spectra are also recorded into their aliased replica fringes.

The numerical analysis for describing these phenomena is shown in Fig. 3. 
A digital hologram of a point object was synthesized using the specifications as follows: the object and hologram spaces consisting of 512$\times$512 pixels with a 8-$\rm{\mu}$m pixel pitch, 
a plane wave with a wavelength of 532 nm, and a critical distance $z_c$ of 61.6 mm. 
The final phase hologram made at a distance of one-third of $z_c$ forms nine replica zones, and the 3-fold expanded angular spectrum was directly calculated.
The Wigner space graphically describes this phenomenon, in Figs. 3(c) and (d). 
The Fresnel transform is represented as the shearing of the Wigner distribution function, $W(x,f_x)=W(x-\lambda zf_x,f_x)$. 
The overlapping of high-order spectra is inevitable at a distance below $z_c$, which causes the aliasing fringe of the hologram having a maximum spatial frequency lower than that of original.
The Wigner distribution is rotated by $90^{\circ}$ via the Fourier transform, which also represents the replications of the angular spectrum.
We find that the aliased replica components of angular spectrum correspond to the replication elements of digital hologram.

The cropped high-frequency components of the digital hologram can be restored by expanding a bandlimited transfer function in the angular spectrum space,
where its expansion process is carried out from its replication during optimization procedure.
Finally, above properties of optical kernel functions lead to the revival of the high-frequency components in the enhanced-NA Fresnel hologram. 
Furthermore, the higher order terms in Eq. (11) can be directly reinterpreted as high-order diffractions in the wave-based propagation model.
The high-order diffractions act as high-frequency components of the diffracted wave, indicating that in a digitized imaging system,
the resolution performance would be enhanced only if high-order diffractions are used.

\begin{figure}[ht!]
\includegraphics[scale=0.73, trim= 0cm 13.7cm 0cm 0cm]{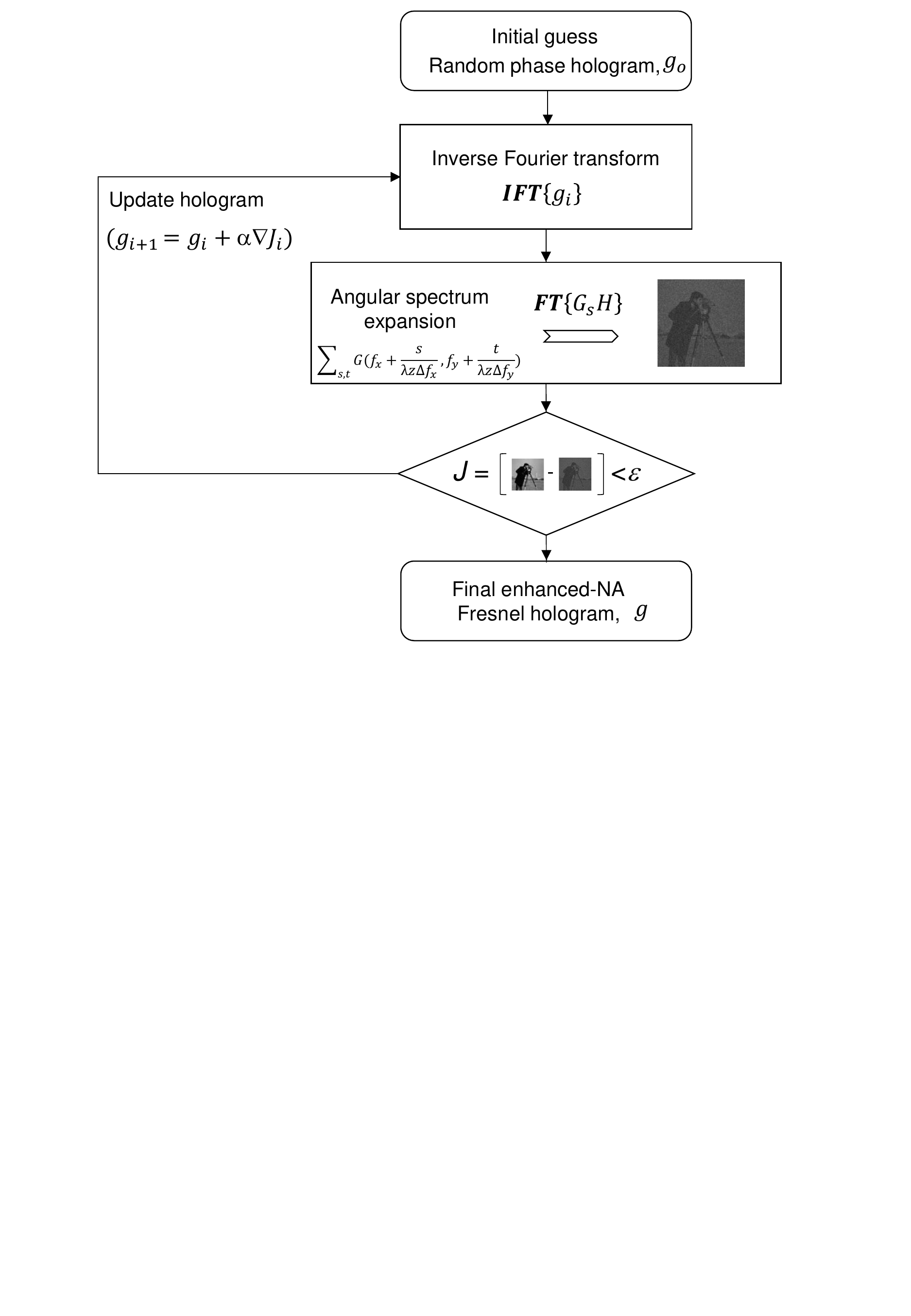}
\caption{Flowchart of the proposed algorithm for optimizing the enhanced-NA Fresnel hologram to recover the low space-bandwidth.
$G_s$ is the expanded angular spectrum multiplied by a sinc function $S$, and $\alpha$ is a learning rate.}
\end{figure}

\section{Numerical results for viewing-angle expansion of holographic image}

We developed the optimization algorithm to expand the angular spectrum of enhanced-NA Fresnel hologram, which is based on gradient descent method, as depicted in Fig. 4.
The gradient descent optimization has been known to show a great performance in hologram synthesis such as neural holography \cite{20}.
The complex-valued hologram $g$ at respective iteration is updated by minimizing the loss function $J$,
 i.e. the squared error between the original image and the reconstructed image:
\begin{equation}
\rm{arg}\min_{\it{g}} \frac{\rm{1}}{2} \| \textbf{\textit{A}}\it{g}-o\| _{\rm{2}} ^{\rm{2}},  \qquad  \textrm{s.t.} \quad G_s=GS.
\end{equation}
The image reconstruction is performed by the inverse transform matrix $\textbf{\textit{A}}$ of Eq. (6), where to revive the high frequency components,
the angular spectrum $G$ of the digital hologram was expanded via its repetition process of $m$ multiples.
The angular spectrum $G_s$ multiplied by the sinc function $S$ caused by a pixelated structure was used, which acts as a constraint in this algorithm.
The corresponding transfer function was also extended to $m$-multiple frequency space.
Finally, the reproduced image undergoing $m$-fold increase in spatial resolution was downsampled to match the original resolution with no averaging process for pixels.

\begin{figure}[ht!]
\includegraphics[scale=0.6, trim= -1.5cm 14.7cm 0cm 0cm]{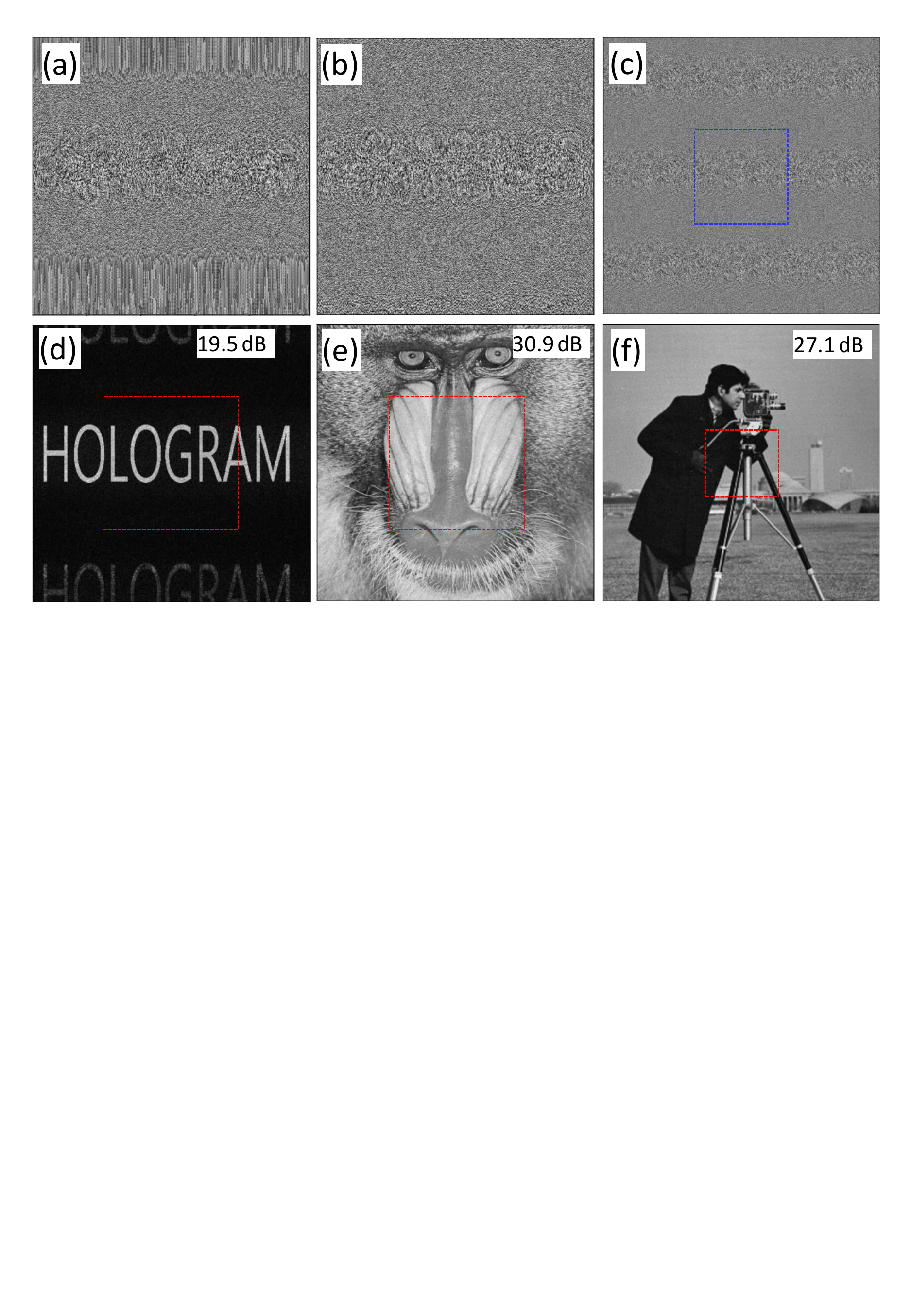}
\caption{Simulation results of the proposed algorithm. Digital holograms made via the processes (a) without and (b) with the angular spectrum expansion.
(c) The angular spectrum in green box is placed at a repetitive interval. (d) Sparse letter image is reconstructed from the hologram synthesized at half-$z_c$. 
Reproduced dense images from the hologram made at (e) half-$z_c$ and (f) quarter-$z_c$, respectively. Red boxes indicate the diffraction zone by a hologram pixel pitch.}
\end{figure}

Figure 5 shows the simulation results obtained based on the proposed scheme.
An object image consisting of 512$\times$512 pixels with an 8-$\mu$m pixel pitch and the plane wave with a wavelength of 532 nm were used.
The critical distance $z_c$ for this specification is computed to be 61.6 mm.
The directly calculated Fresnel hologram using the “HOLOGRAM” letter image at a distance of half of $z_c$ shows 
that the aliased replica fringes are cropped in the vertical direction because of the use of only the zeroth-order region in the angular
spectrum space, as illustrated in Fig. 5(a). 
The lateral size of the active area in the hologram is two times smaller than the original size.
The value decreases in proportion to the located distance of the object image.
Based on Eq. (4), the hNA is constant irrespective of the distance, thus preventing the increase of the viewing angle.
The viewing angle $\it{\Omega}$ of the reconstructed image has the same quantity as the diffraction angle $\theta$ of 3.8$^\circ$.

Figure 5(b) is the digital hologram synthesized by using the proposed algorithm.
The angular spectrum was expanded to 3 multiples in the intermediate process.
The aliased fringes in the angular spectrum appear in the seamless form among the replicas after optimization, in Fig. 5(c).
We find that the active region of the hologram field occupies the entire area of the digital hologram, 
which was clearly confirmed on a logarithmic scale.
The digital hologram made at a distance of half $z_c$ undergoes two-fold increase in the hNA.
Further, we confirmed this expansion of the hologram field in the synthesized hologram at a distance of quarter $z_c$.
Figure 5(d) is the reproduced image through the algorithm, which becomes a standard for algorithm performance.
Here, the simulated phase hologram was used.
The high-order images in the vertical direction are not completely removed.
On the other hand, the dense image occupying the entire area of the object space is well reconstructed without obstructing high-order images, 
as shown in Figs. 5(e) and 5(f).
The PSNR values of the reproduced images from the holograms placed at half-$z_c$ and quarter-$z_c$ distances are estimated 
to be approximately 30.9 dB and 27.1 dB, respectively.
The angular spectrum of a digital hologram made at a quarter-$z_c$ distance was extended to a 5-multiple scope to cover sufficient high-frequency components.

\begin{figure}[ht!]
\includegraphics[scale=0.8, trim= 1cm 16cm 0cm 0cm]{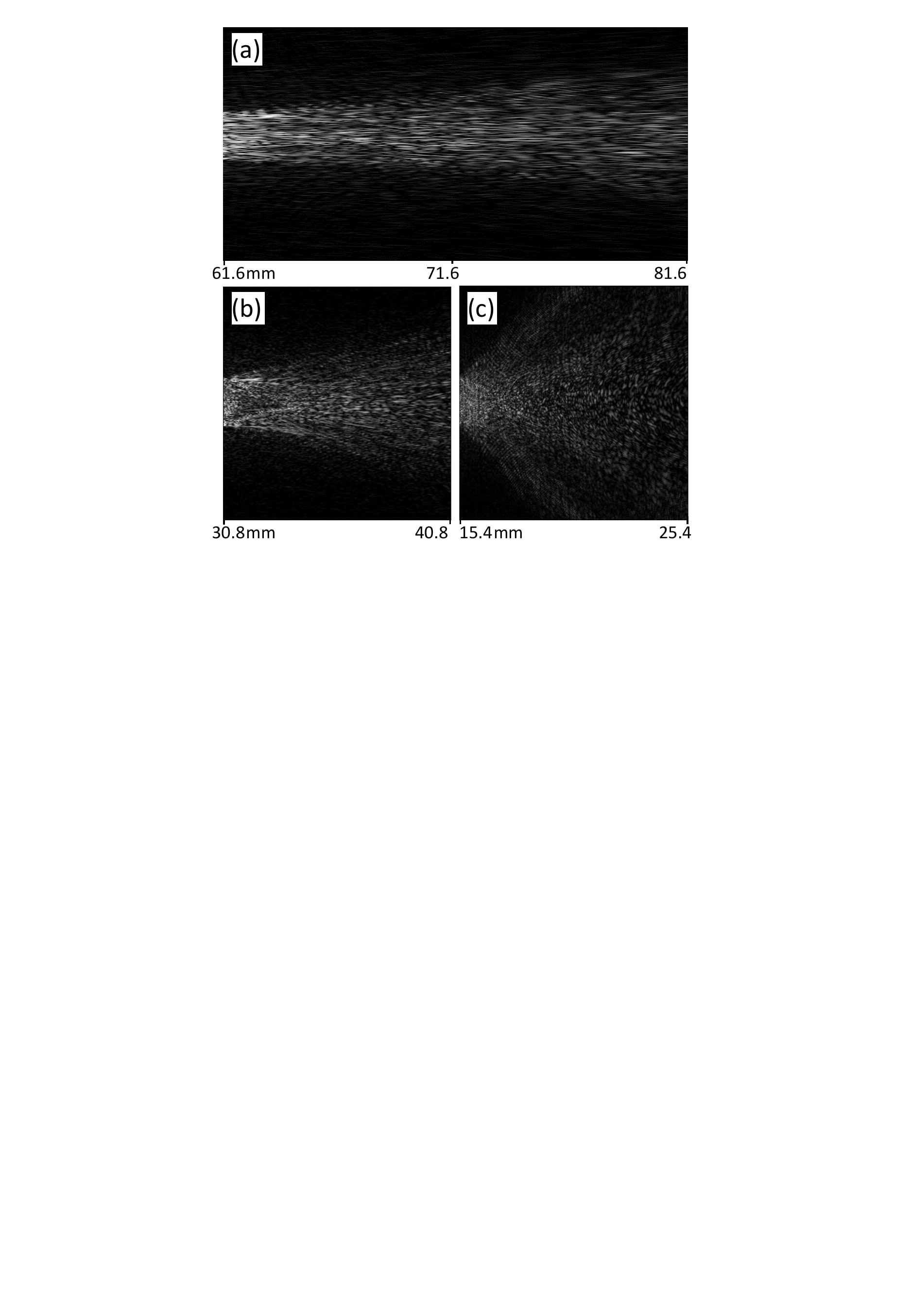}
\caption{Diffractive waves propagating away from the focused images.
The focused images are generated from the holograms synthesized at (a) a critical distance $z_c$, (b) half-$z_c$, and (c) quarter-$z_c$, respectively.
Lateral sizes of the corresponding images are 4096 $\mu$m, 2048 $\mu$m, and 1024 $\mu$m, respectively.}
\end{figure}

Figure 6 illustrates the diffractive waves that propagate away from the focused images.
The phase hologram is taken from the complex modulation because it accelerates convergence in the algorithm, 
even when using a severely sparse object.
Measuring the spreading angle of the diffractive wave is robust method for confirming the viewing angle \cite{17}. 
The Fresnel transform with a single Fourier transform based on Eq. (2) was used to exclude an 
artificial expansion of the angular spectrum because it confines an angular view.
The intensity of a particular vertical line in the letter image is displayed with a propagating distance.
We can observe that the spreading angle increases with decreasing synthesis distance of the digital hologram,
indicating that the viewing angle is larger at a shorter synthesis distance.
The viewing angle of the reconstructed image from the digital hologram made at a critical distance is estimated to be about 3.7$^\circ$.
The reconstructed images at half-$z_c$ and quarter-$z_c$ distances reveal viewing angles of approximately 7.0$^\circ$ and 13.1$^\circ$, respectively.
It is confirmed that the proposed algorithm restores the images at a viewing angle dependent on the hNA.

\begin{figure}[ht!]
\includegraphics[scale=0.75, trim= 1cm 12.5cm 0cm 0cm]{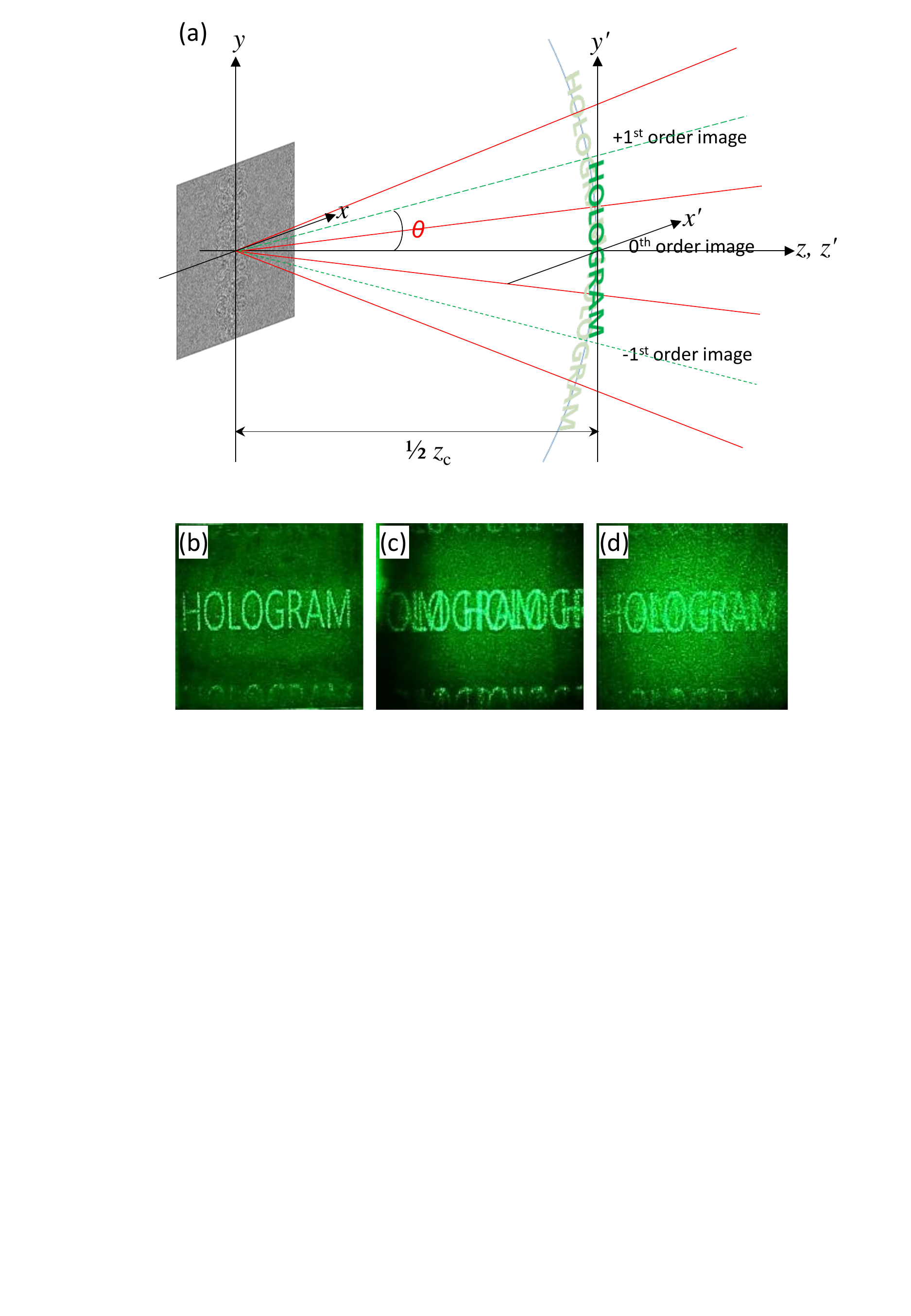}
\caption{(a) Configuration of image reconstruction from enhanced-NA Fresnel hologram.
For clarity, only three images, i.e., those of $0^{\rm{th}}$- and $\pm 1^{\rm{st}}$-orders are depicted in vertical direction.
(b) Optically captured images in the viewing direction of the zeroth order. 
Captured images in the slightly shifted direction into the first-order viewing zone (c) without and (d) with the angular spatial filter (see {\color{blue} Visualization 1}).}
\end{figure}

\section{Optical experiments and discussions}

Optical experiments were carried out by using a phase-only SLM (Holoeye PLUTO) with 1920$\times$1080 pixels of an 8-$\mu$m pixel pitch and 
a green laser with a wavelength of 532 nm.
A digital hologram for the letter image with 1024$\times$1024 pixels was prepared, in which the zero values were filled outside the bounds.
The hologram was made at half-$z_c$, 61.6 mm.
Figure 7(a) shows the configuration of image reconstruction on the direction cosine surface.
The complete image without an intefernce of high-order terms is well reconstructed in the central direction,
because only the image in the plane vertical to the central axis was considered during the optimization algorithm \cite{17,21}.
However, when the viewing direction deviates from the central axis, the first-order image appears, and overlaps with the zeroth-order image,
because the pixelated structure of SLM creates the high-order images.

We focused two points in this optical experiment.
First, optical reconstruction behaves in accordance with the simulation results.
Second, the high-order images are carried by corresponding diffraction waves in an oblique view direction.
In this situation, it provides the possiblity to remove separately the high-order noises using an appropriate spatial fitering technique.  

As displayed in Fig. 7(b),
the hologram reconstructs the original letter image with no overlapping of high-order noises.
The polarization filtering technique was used to eliminate the non-diffracted beams. 
Noise remnants exist in the vertical direction due to the vacant space in a sparse image.
When the view direction angularly moves the first-order viewing zone, the zeroth-order image is still viewable.
Therefore, the viewing angle is twice as large as the diffraction angle.

When the viewing direction slightly deviates from the central axis, an overlapping with the first-order image occurs, in Fig. 7(c).
Recently, it was reported that an angular optical filter can block high-order diffractive beams \cite{22}.
The used optical filter was a bandpass filter (532 nm MaxLine laser clean-up filter, Semrock) with extrmely low peak width of 2 nm,
where angluar range to transmit a light is about 11$^\circ$.
To overcome a relatively wide angular range in comparision to a diffraction angle, 
we carried out the blocking experiment by tilting the optical filter with respect to normal direction,
resulting in the observation of this phenomenon even at a few degrees.
Figure 7(d) is the captured image through this technique.
The suppression of the first-order image clearly appears in the slightly shifted viewing-direction.
This indicates that the first-order image is carried by an obliquely diffractive wave.
The blockade was lifted at the largely shifted position, based on Fig. 7(a).
We know that if optical properties are different from those of the zeroth-order beam, high-order noises could be removed using an spatial filter.

Meanwhile, the holographic near-eye display for virtual and augmented reality requires a sufficient eyebox to accomplish the large field of view of holographic image \cite{23}.
In a near-eye display system, the viewing direction does not change while seeing a floating image, and thus it can avoid the obstruction of high-order noises.
The expanded viewing-angle based on the enhanced-NA Fresnel hologram could contribute to the eyebox expansion in the holographic near-eye display.

\section{Conclusions}

In a digital hologram with a finite bandwidth, it is a hard task in generating a holographic image at a viewing angle higher than the diffraction angle without sacrificing an image size.
The formation of replica fringes acting as an origin of overlapping with high-order images should be prevented above all, and furthermore the hNA of initial system has to be secured.
We present an effective method for synthesizing  the Fresnel hologram with a high numerical aperture by using the algorithm consisting of the double Fourier transform. 
The expansion of the angular spectrum in Fourier space leads to the viewing angle dependent on the numerical aperture of the digital hologram. 
However, the proposed algorithm is limited to removing high-order diffraction images because the pixelated SLM creates optically high-order diffractive waves.
We found that through the experiment eliminating these terms by an angular spatial filter, the high-order images are carried by diffractive waves in an inclined angle.
When the diffractive waves has a different optical properties such as a polarization state, a complete removal of them comes true by an spatial filter. 
Further study to block high-order noises will be progressed in a future research.
This approach offers the posibility to expand a viewing angle of holographic image even using the conventional spatial light modulator.

\begin{backmatter}
\bmsection{Funding}
This work was partially supported by Institute for Information \& Communications Technology Promotion (IITP) grant funded by the Korea government (MSIP) (2017-0-00049 and 2021-0-00745)

\bmsection{Disclosures}
The authors declare no conflicts of interest.

\bmsection{Data Availability Statement}
Data underlying the results presented in this paper are not publicly available at this time but may be obtained from the authors upon reasonable request.

\end{backmatter}

\end{document}